\documentclass[usenatbib,usegraphicx]{mn2e}
\usepackage{amssymb}
\usepackage{color}
\usepackage{graphicx}

\newcommand{\halfspace}{\hspace{1pt}}

\newcommand{\MSun}{\mathop{\rm M_\odot}\nolimits}

\newcommand{\kms}{\mathop{\rm km \ s^{-1}\,}\nolimits}
\newcommand{\Lya}{Ly$\alpha$}

\newcommand\HI{{\hbox{H\halfspace$\rm \scriptstyle I$}}}

\newcommand\lsim{~\lower.5ex\hbox{$\buildrel < \over \sim$}~}
\newcommand\gsim{~\lower.5ex\hbox{$\buildrel > \over \sim$}~}

\title[Gas around galaxies III:\ hydrogen absorption]{Gas
  around galaxy haloes - III: hydrogen absorption signatures around
  galaxies and QSOs in the Sherwood simulation suite}
       
\author[Avery Meiksin, James S. Bolton, Ewald Puchwein]{
        Avery Meiksin$^{1}$\thanks{E-mail:\ A.Meiksin@ed.ac.uk (AM)},
        James S. Bolton$^{2}$, Ewald Puchwein$^{3}$\\
        $^{1}$SUPA\thanks{Scottish Universities Physics Alliance},
	Institute for Astronomy, University of Edinburgh,
        Blackford Hill, Edinburgh\ EH9\ 3HJ, UK\\
        $^{2}$School of Physics and Astronomy, University of Nottingham,
        University Park, Nottingham\ NG7\ 2RD, UK\\
        $^{3}$Kavli Institute for Cosmology and Institute of Astronomy,
        Madingley Road, Cambridge, CB3\ 0HA, UK}

\shortcites{}

\begin{document}

\date{Accepted . Received ; in original form }
\pagerange{\pageref{firstpage}--\pageref{lastpage}} \pubyear{2016}
\maketitle
\label{firstpage}

\begin{abstract}

  Modern theories of galaxy formation predict that galaxies impact on
  their gaseous surroundings, playing the fundamental role of
  regulating the amount of gas converted into stars. While
  star-forming galaxies are believed to provide feedback through
  galactic winds, Quasi-Stellar Objects (QSOs) are believed instead to
  provide feedback through the heat generated by accretion onto a
  central supermassive black hole. A quantitative difference in the
  impact of feedback on the gaseous environments of star-forming
  galaxies and QSOs has not been established through direct
  observations. Using the Sherwood cosmological simulations, we
  demonstrate that measurements of neutral hydrogen in the vicinity of
  star-forming galaxies and QSOs during the era of peak galaxy
  formation show excess Lyman-$\alpha$ absorption extending up to
  comoving radii of $\sim150$~kpc for star-forming galaxies and
  $300-700$~kpc for QSOs. Simulations including supernovae-driven
  winds account for the absorption around star-forming galaxies but
  not QSOs.

\end{abstract}

\begin{keywords}
cosmology:\ large-scale structure of Universe -- quasars: absorption lines --
galaxies:\ formation -- intergalactic medium
\end{keywords}

\section{Introduction}
\label{sec:Intro}

The paradigm of cosmological structure formation through the growth of
primordial density fluctuations predicts galaxies reside within
accreting haloes at the interstices of a cosmic web of dark matter and
gas. The most straightforward estimates of the expected stellar
content of galaxies, however, suggest a conversion rate of infalling
gas into stars that is too high. Galaxies should be much brighter than
measured. It is widely believed some form of feedback must regulate
the efficiency of star formation. Both semi-analytic models and
cosmological hydrodynamics simulations suggest two forms of feedback
are in play:\ stellar and supernovae-driven galactic winds in
star-forming galaxies and heating generated through accretion onto a
supermassive black hole in galaxies with an Active Galactic Nucleus
(AGN) \citep[see the review by][]{2015ARAA..53...51S}. The estimated
galactic stellar mass dividing star-forming galaxies from the most
massive AGN, Quasi-Stellar Objects (QSOs), corresponds to a dark
matter halo mass of about $10^{12}\MSun$ \citep{2003MNRAS.346.1055K,
  2012ApJ...752...39T}. This is the halo mass for which the efficiency
of gas conversion into stars peaks \citep{2013ApJ...770...57B}.

The opportunity to detect direct evidence for any difference in the
environmental impact between star-forming galaxies and QSOs, however,
has been wanting. This has now changed with the advent of
observational programmes that probe the gaseous surroundings of
galaxies and QSOs. Utilizing the large-scale QSO surveys from the
Sloan Digital Sky Survey \citep{2000AJ....120.1579Y,
  2005AJ....130..367S} and its extension to the Baryon Oscillation
Spectroscopic Survey \citep{2004MNRAS.349.1397C}, the
Quasar-Probing-Quasar survey (QPQ) \citep{2006ApJ...651...61H} has
used lines of sight to 149 QSOs to probe the gaseous surroundings of
both foreground QSOs \citep{2009ApJ...690.1558P, 2013ApJ...776..136P}
and galaxies \citep{2013ApJ...762L..19P}. Similarly, the Keck Baryonic
Structure Survey (KBSS) \citep{2012ApJ...751...94R,
  2012ApJ...750...67R, 2012ApJ...752...39T} has used lines of sight to
15 hyperluminous QSOs to measure gaseous absorption through the
extended haloes of more than a thousand galaxies. Interest has
particularly focussed on the central few hundred kiloparsecs, the
Circumgalactic Medium (CGM), since this is the arena in which
supernovae-driven winds in star-forming galaxies and heating driven by
black hole accretion in QSOs are expected to have their greatest
impact \citep[e.g.][]{2002MNRAS.330..113K, 2010MNRAS.402.1536S,
  2011MNRAS.415.2782V, 2013MNRAS.435.2931H}.

The QPQ and KBSS results demonstrate galaxies and QSOs are surrounded
by excess amounts of Lyman-$\alpha$ absorption over that arising from
the large-scale distribution of gas, the Intergalactic Medium (IGM)
\citep{2006ApJ...651...61H, 2010ApJ...717..289S, 2012ApJ...751...94R,
  2013ApJ...776..136P}. Measuring the departures from the absorption
predictions of models that exclude feedback provides a novel means of
quantifying the impact of the winds and black hole accretion. Matching
the excess absorption provides a test of feedback models. Accounting
for the covering fractions of neutral hydrogen absorption systems,
especially those optically thick to photoionizing radiation, has
proven particularly challenging. Motivated by the search for evidence
of cold streams feeding the growth of galaxies
\citep{2003MNRAS.345..349B, 2005MNRAS.363....2K}, early simulations
produced substantially smaller covering fractions in massive haloes
than measured. More recent simulations have had greater success, but
still have not fully accounted for the observations. Using the FIRE
\lq\lq zoom-in\rq\rq simulations \citep{2014MNRAS.445..581H}, which
include feedback from stellar winds and supernovae but not from AGN,
\citet{2015MNRAS.449..987F} reproduced the average covering fraction
of Lyman Limit Systems, absorbers optically thick at the Lyman edge
($N_{\rm HI}>10^{17.2}\,{\rm cm^{-2}}$), for lines of sight passing
within the virial radius of star-forming galaxies, as measured by
\citet{2012ApJ...750...67R}, but not for the QSO measurements of
\citet{2013ApJ...776..136P}. The model prediction for star-forming
galaxies for the average covering fraction within the virial radius of
absorption systems with $N_{\rm HI}>10^{15.5}\,{\rm cm^{-2}}$ was a
factor $\sim2$ short of measurements. On increasing the mass
resolution, \citet{2016MNRAS.461L..32F} were able to reproduce the
covering fraction of optically thick absorbers within the virial
radius of QSOs. Notably AGN feedback was still not included. Using a
different zoom-in simulation with feedback from stellar winds and
supernovae, but not AGN \citep[][and references
  therein]{2013MNRAS.435..999D}, \citet{2014ApJ...780...74F} found a
covering fraction for optically thick absorbers within the virial
radius of star-forming galaxies about half that measured, but
consistent within the broad observational error. The prediction for
QSOs fell well short of measurements. The dark matter particle mass
was only twice that of the higher resolution FIRE simulations used,
but the baryon mass resolution achieved may have been considerably
poorer. The simulations also over-predict the stellar content of the
galaxies by a factor of two, so that the gas content was likely too
low. Using a suite of cosmological hydrodynamics simulations with a
variety of wind feedback models, both with and without AGN feedback,
\citet{2015MNRAS.448..895S} found some of the models match the
cumulative covering fraction of optically thick absorbers within the
virial radius of star-forming galaxies, but the predicted values are
about half the measured value within twice the virial radius. In an
analysis of the EAGLE simulations \citep{2015MNRAS.446..521Sfull}, for
which a different supernova-driven wind feedback model was implemented
and which include AGN heating, \citet{2015MNRAS.452.2034R} matched the
covering fraction profile of systems with column densities exceeding
$10^{15.5}\,{\rm cm^{-2}}$ around star-forming galaxies for lines of
sight with a range of impact parameters both inside and outside the
virial radii. They were similarly able to reproduce the profiles for
the covering fractions of optically thick systems around QSOs. As
noted by \citet{2016MNRAS.461L..32F}, however, the covering fraction
of optically thick absorbers increases rapidly with halo mass and
\citet{2015MNRAS.452.2034R} confined their comparison to QSO halo
virial masses $M_h>10^{12.5}\,\MSun$. Observations favour a somewhat
lower mass range \citep{2012ApJ...752...39T, 2012MNRAS.424..933W,
  2015MNRAS.453.2779E}. Since the simulation used included AGN
feedback, it is also not known whether supernovae-driven winds would
have been sufficient. The mass resolution of the simulation was
moreover intermediate between the FIRE simulations that succeeded in
reproducing the covering fraction around QSOs without AGN feedback and
those that failed, so it is not known if this is a critical factor as
well. On the other hand, an advantage of cosmological simulations over
zoom-in simulations is the large sample size, with over a hundred QSO
halo mass systems at $z=2$ in the EAGLE simulation, which may be more
representative of the population than the $\sim15$ systems available
in the FIRE simulation, and fewer in the \citet{2014ApJ...780...74F}
analysis.

The covering fractions do not provide a full description of hydrogen
absorption around galaxies and QSOs. The collective absorption may be
quantified additionally by the integrated equivalent width of
absorption along the line of sight, or, analogously, the excess in the
mean absorbed flux over the contribution from the diffuse IGM. Using a
zoom-in simulation of a single galaxy including supernovae-driven wind
feedback, \citet{2013ApJ...765...89S} matched the equivalent widths
measured outside the virial radii of star-forming galaxies in the KBSS
survey \citep{2010ApJ...717..289S, 2012ApJ...751...94R}, but the
predictions fell nearly a factor of two short for lines of sight with
impact parameter at about half the virial radii, not much different
from the case with no feedback \citep{2015MNRAS.453..899M}. The
simulations successfully reproduced the mean covering fraction of
optically thick absorbers within the virial radii of star-forming
galaxies, but were a factor of two low at twice the virial radius. The
covering fraction for systems with $N_{\rm HI}>10^{15.5}\,{\rm
  cm^{-2}}$ was also low compared with the measured values. The virial
mass of the galaxy halo was $2.6\times10^{11}\,M_\odot$, somewhat low
for the population of star-forming galaxies in the KBSS survey, but
the results illustrate that a variety of absorption measurements,
including profile information, must be used to demonstrate that a
simulation model gives an accurate representation of the distribution
of neutral hydrogen around galaxies.

A principal goal of this paper is to establish the scale of departure
of the gaseous haloes around galaxies from the non-feedback cosmic-web
prediction. Previously we found models without feedback matched the
observed absorption signatures of the \lq\lq mesogalactic
medium\rq\rq~at distances exceeding $500-800$ comoving kpc around
galaxies with halo masses up to $10^{12}\,\MSun$ at redshifts
$2<z<2.5$ \citep{2015MNRAS.453..899M}; the simulation volumes were too
small to make firm statements for larger mass haloes. By contrast,
clear excesses in the data compared with the non-feedback models were
found within the inner $100-300$ comoving kpc
\citep{2015MNRAS.453..899M}. Using the Sherwood simulation suite
\citep{2016arXiv160503462B}, we reinvestigate the signatures of
hydrogen absorption around galaxies and QSOs. We include simulations
both without and with feedback, using a different feedback model
compared with previous simulations in the context of CGM measurements
to further test the sensitivity of the absorption signatures to the
feedback model. These new larger simulations also substantially
improve the precision of the predictions, particularly for the rare
massive haloes exceeding $10^{12}\,\MSun$, enabling us to quantify the
differences between the absorption signatures below and above this
mass threshold.

This paper is organized as follows. In Sec.~\ref{sec:sims} we
summarise the numerical simulations used in this work. The resulting
\HI\ absorption signatures are presented in Sec.~\ref{sec:results}. We
summarize and discuss our conclusions in
Sec.~\ref{sec:conclusions}. Simulation convergence tests are presented
in an Appendix. Unless stated otherwise, all results are for a flat
$\Lambda$CDM universe with the cosmological parameters
$\Omega_m=0.308$, $\Omega_bh^2=0.0222$ and $h=H_0/100~\kms\,{\rm
  Mpc}^{-1}=0.678$, representing the present-day total mass density,
baryon density and Hubble constant, respectively. The power spectrum
has spectral index $n_{\rm s}=0.961$, and is normalized to
$\sigma_{8}=0.829$, consistent with the Cosmic Microwave Background
data from {\it Planck} \citep{2014A&A...571A..16P}. All distances are
comoving unless stated otherwise; \lq cMpc\rq is used to designate
comoving megaparsecs, and \lq ckpc\rq to designate comoving
kiloparsecs.

\section{Numerical simulations}
\label{sec:sims}
 
\begin{table*}
  \centering  
    \begin{minipage}{180mm}
      \begin{center}
        \caption{Summary of the simulations performed in this work. 
          The columns, from left to right, list the simulation name,
          the comoving box size, the dark matter particle mass,
          the baryon particle mass, the comoving gravitational softening length
          and the method of gas removal/ wind feedback.}
    \begin{tabular}{l c c c c r}
      \hline\hline  
      Name       & Box size     & M$_{\rm dm}$ & M$_{\rm gas}$    & $l_{\rm soft}$  
      & gas removal/ wind       \\
                 & [$h^{-1}$~cMpc] & [$h^{-1}\,\MSun$] & [$h^{-1}\,\MSun$] &
                                                                   [$h^{-1}$~ckpc] \\
      \hline  
      40-2048  &      40  &  $5.37\times10^5$ &  $9.97\times10^4$  &
      0.78 & qLy$\alpha$ \\
      80-2048  &      80  &  $4.30\times10^6$ &  $7.97\times10^5$  &
      1.56 & qLy$\alpha$ \\
      40-1024  &      40  &  $4.30\times10^6$ &  $7.97\times10^5$  &
      1.56 & qLy$\alpha$ \\
      40-1024-ps13  &      40  &  $4.30\times10^6$ &  $7.97\times10^5$  &
      1.56 & PS13 \\
      20-512  &      20  &  $4.30\times10^6$ &  $7.97\times10^5$  &
      1.56 & qLy$\alpha$ \\
      \hline  
    \end{tabular}
    \label{tab:sims}
\end{center}
  \end{minipage}
\end{table*}
 
The numerical simulations are performed using a modified version of
the parallel Tree-PM SPH code \texttt{P-GADGET-3}, an updated version
of the publicly available code \texttt{GADGET-2} \citep[last described
  by][]{2005MNRAS.364.1105S}. Both simulations without and with
galactic winds were performed. Because the computational demands for
generating a galactic wind from first principles are well beyond
current resources, all wind feedback models rely on a subgrid
approximation scheme. Previously \citep{2015MNRAS.453..899M}, we had
used a star-formation and feedback model with a constant wind velocity
for all haloes \citep{2003MNRAS.339..289S}. Such models, however, do
not well reproduce the luminosity function of galaxies. The wind
simulation used here is based on the star-formation and wind model of
\citet[PS13; ][]{2013MNRAS.428.2966P}. This scheme is built on a
subgrid model that allows for star formation in a multiphase
interstellar medium, with the feedback provided by stochastic kinetic
energy injection by assigning to particles entering the wind near the
centre of a galaxy a kick velocity that scales with the escape
velocity from the halo. The mass loading factor, defined as the ratio
of the mass flow in the wind to the star formation rate, is
proportional to the inverse square of the escape velocity, thus using
a constant amount of energy per unit mass of stars formed for driving
the wind. This form has been found to recover the luminosity function
and luminosity-metallicity relation of Local Group satellite galaxies
\citep{2010MNRAS.406..208O}. It also reproduces the stellar mass
function of galaxies over redshifts $z<2$, the observed gas to stellar
mass ratios and specific star formation rates as a function of stellar
mass \citep{2013MNRAS.428.2966P}, and a variety of other galaxy
properties over a wide range of redshifts
\citep{2013MNRAS.436.3031V}. It receives additional support from x-ray
measurements. The measured linear proportionality of the diffuse soft
x-ray luminosity of galaxies with the star formation rate
\citep{2012MNRAS.426.1870M}, is matched in an energy-driven wind model
for an asymptotic wind velocity that scales approximately as $\dot
M_*^{1/6}$ \citep{2016MNRAS.461.2762M}. For a star-formation rate
scaling approximately like the stellar mass
\citep[e.g.][]{2007AA...468...33E} and the ratio of stellar mass to
halo mass scaling approximately as the halo mass for halo masses
$10^{11}<M_h<10^{12}\,M_\odot$ \citep[e.g.][]{2013ApJ...770...57B},
the scaling required by the x-ray measurements corresponds at least
approximately to a wind velocity scaling like the halo escape
velocity. The mass-loading factor for the wind model in the
simulations has typical values of $1-10$. This is consistent with the
values found for the FIRE simulations \citep{2015MNRAS.454.2691M}. By
contrast, the mass loading factor was less than unity for the galaxy
in \citet{2013ApJ...765...89S}. Mass loading factors have not been
published for the EAGLE simulations.
 
Other feedback mechanisms are considered in the literature of galaxy
formation simulations. \citet{2005ApJ...618..569M} consider winds
driven by momentum-deposition from radiation pressure or
supernovae. \citet{2014MNRAS.442.3013K} use a superbubble feedback
model for which mass loading includes a contribution from thermal
evaporation. Another possibly important wind driver is cosmic-ray
pressure \citep{1975ApJ...196..107I}, which has recently been used in
cosmological galaxy formation simulations \citep{2014ApJ...797L..18S,
  2016ApJ...824L..30P}.

The computational methods are described in detail in
\citet{2016arXiv160503462B}. The runs used in this paper are
summarized in Table~\ref{tab:sims}. Except for the wind simulation,
the computations use the \lq\lq quick Ly$\alpha$\rq\rq method
(qLy$\alpha$), for which all gas cooler than $10^5$~K and with an
overdensity exceeding 1000 is converted into collisionless \lq\lq
star\rq\rq particles (without feedback). This is a computational trick
used in IGM simulations to significantly speed up the computation
\citep{2004MNRAS.354..684V}, and is not meant to represent the stellar
content of real galaxies. A detailed comparison with similar
simulations using the finite-difference code \texttt{Enzo}
\citep{2014ApJS..211...19B}, with no gas removal or feedback
implemented, shows that, while the quick Ly$\alpha$ approximation
reduces the circumgalactic gas density within the virial radius of
haloes compared with \texttt{Enzo}, the integrated absorption
signatures are unaffected along baselines comparable to those used in
measurements \citep{2015MNRAS.453..899M}, even for lines of sight
passing within the virial radius. This was shown to be because the
integrated absorption signals over baselines exceeding $\sim1000\kms$
are dominated by gas outside the virial radius for non-feedback models
even for impact parameters smaller than the virial radius. We
accordingly use the quick Ly$\alpha$ simulations to model the
absorption signatures in the absence of feedback.

To interpret the large-scale mesogalactic absorption signal through
the extended galactic haloes outside the circumgalactic region, we
also compute simple one-dimensional spherically symmetric collapse
models. We use the combined gravitating shell model and hydrodynamics
finite difference code described in \citet{Meiksin94}, modified to
allow for a cosmology with a cosmological constant. To recover the
asymptotic mean profile at large distances from the centre, we use the
mean profile derived by \citet{1986ApJ...304...15B} as the initial
density profile. Since the gas is smooth, it is necessary to adopt a
method to represent the density fluctuations underlying the Ly$\alpha$
forest, without which the important effects of line blanketing will
not be recovered. We do so by implementing the lognormal density
fluctuation scheme of \citet{1993ApJ...405..479B} and
\citet{1997ApJ...479..523B} along projected one-dimensional lines of
sight through the spherically symmetric gas distribution. Details are
described in \citet{2014MNRAS.437.3639C}. To describe the fluctuations
around a halo, the density field is normalized to the local mean
density in the halo.

For both the full cosmological simulation and the spherically
symmetric code, atomic radiative cooling from hydrogen and helium are
included, along with photoionization heating. Metal-line cooling is
not included. Metal-line cooling negligibly affects the large-scale
IGM, but may play an important role within galactic haloes following
the distribution of metals by supernovae feedback. The gas is
photoionized by a metagalactic ultra-violet background. The mean
Ly$\alpha$ absorbed flux level is set in order to match the
measurements of \citet{2013MNRAS.430.2067B}. Radiative transfer of
continuum radiation is not included in the computations. Radiative
transfer normally has little effect on the temperature of absorption
systems optically thick to hydrogen and helium ionizing photons
because their density is sufficiently high that thermal balance is
achieved. It will, however, affect the amount of hydrogen absorption
in optically thick systems. Because the large velocity shearing of
absorption systems within the virial radius increases the equivalent
widths of the systems \citep{2015MNRAS.453..899M}, especially in the
presence of a wind, it is important to include radiative transfer
effects for predicting the amount of absorption within the virial
radii of the haloes. Ideally, self-consistent radiative hydrodynamics
should be implemented, but this is beyond the scope of this
work. Instead we post-process the ionization fractions using the
approximate treatment of \citet{2013MNRAS.430.2427R}. The effect the
radiative transfer correction has on the absorption signature is
described in the Appendix. The full effects including radiative
hydrodynamics is still a largely unexplored topic.

We extract spectra of the Ly$\alpha$ forest following the standard
procedure, as summarized in \citet{2015MNRAS.453..899M}. A full Voigt
profile is used throughout the analyses. To match the resolution of
the observations, we smooth the resulting spectra with a Gaussian of
FWHM $125\kms$ for comparison with \citet{2013ApJ...776..136P} and
FWHM $8\kms$ for comparison with \citet{2012ApJ...751...94R}. We
extract spectra for a range of impact parameters around haloes
identified in the simulations using the friend-of-friends algorithm,
as described in \citet{2016arXiv160503462B}. We base our comparison
primarily on the Lyman-$\alpha$ absorption equivalent width integrated
across velocity windows. Previously we found this provided an adequate
description of the effect of overdense structures near the haloes on
the absorption statistics \citep{2015MNRAS.453..899M}. For a halo at
measured velocity $v_{\rm halo}$ along a line of sight to a background
QSO, the equivalent width at the Lyman-$\alpha$ wavelength
$\lambda_\alpha$ is computed as
\begin{equation}
  w(b_\perp, \Delta v) = \frac{\lambda_\alpha}{c}\int_{v_{\rm halo}-\Delta v/2}^{v_{\rm halo}+\Delta v/2}\,dv \left[1-e^{-\tau_\alpha(b_\perp, v)}\right],
\label{eq:ewDv}
\end{equation} 
over a velocity window of width $\Delta v$ centred on the position of
the halo, displaced transversely by an amount $b_\perp$. The
equivalent width is then used to define the fractional change
$\delta_F$ in the transmitted flux compared with the mean
intergalactic value for the corresponding velocity
window \citep{2013ApJ...776..136P}
\begin{equation}
\delta_F(b_\perp, \Delta v)=\frac{w(b_\perp, \Delta v) - w_{\rm
    IGM}}{\Delta\lambda - w_{\rm IGM}}.
\label{eq:dFDv}
\end{equation}
Here $\Delta\lambda = \lambda_\alpha\Delta v/c$ and $w_{\rm
  IGM}=\Delta\lambda[1-\exp(-\tau_{\rm eff})]$, where $\tau_{\rm eff}$
is the Lyman-$\alpha$ effective optical depth of the IGM.

In the Appendix we present a suite of convergence tests of the
cosmological simulations, varying the mass resolution and box size. In
brief, we find that a comoving box size of $40h^{-1}$~Mpc with a dark
matter particle mass resolution $4.30\times10^6\,h^{-1}\MSun$
($1024^3$ particles) yields adequate convergence. Further convergence
tests of the absorption signatures are described in
\citet{2015MNRAS.453..899M}.

\section{Results}
\label{sec:results}

\subsection{Deviation of \Lya\ absorption from the mean IGM}

\begin{figure*}
\scalebox{0.5}{\includegraphics{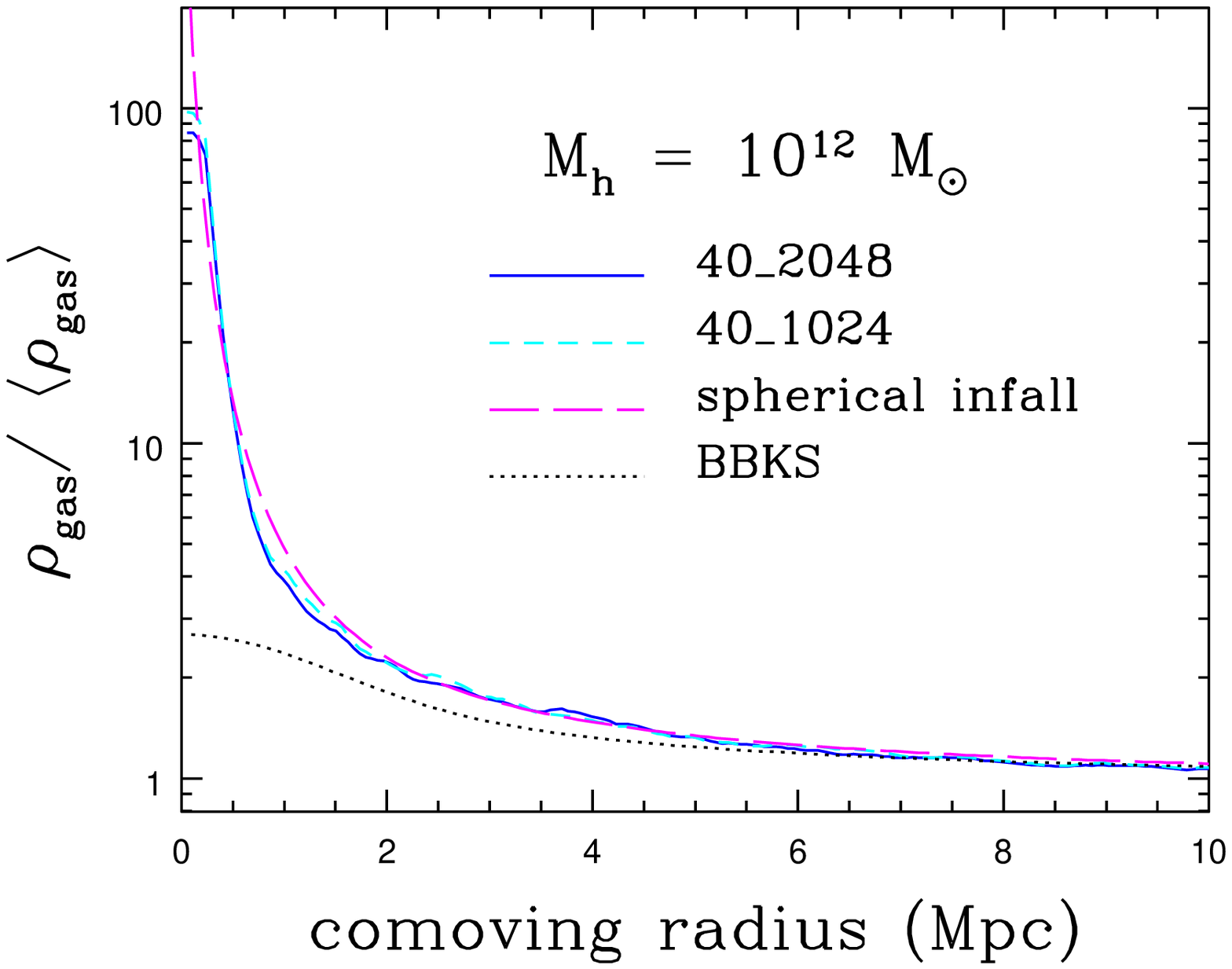}\includegraphics{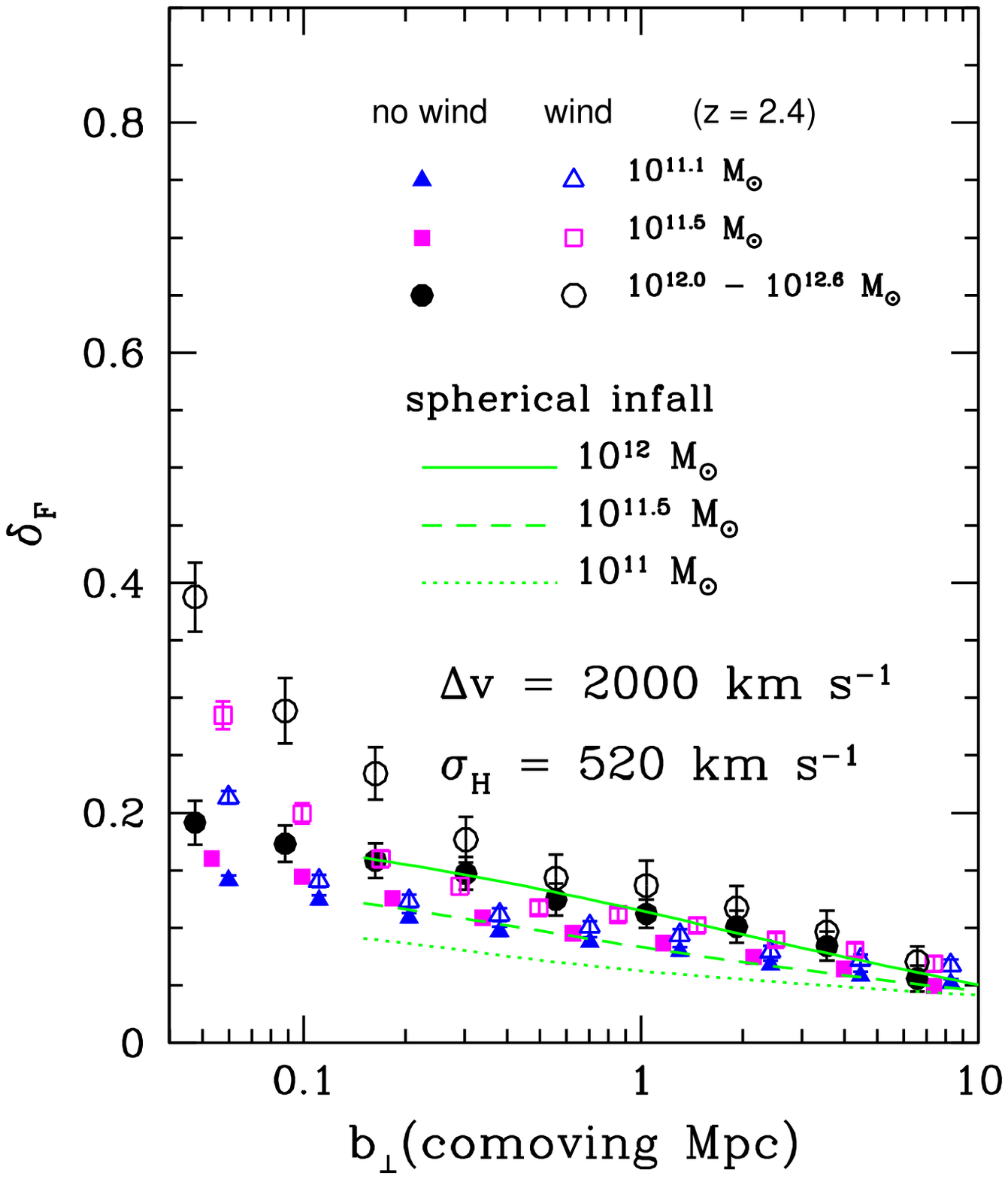}}
\vspace{-1.5cm}
\caption{{\it Left-hand panel:} Radial-averaged gas overdensity
  profiles around haloes with mass $M_h=10^{12}\,\MSun$ at redshift
  $z=2.4$ for non-wind simulations with $2048^3$ dark matter particles
  (40-2048; solid blue) and $1024^3$ dark matter particles (40-1024;
  short-dashed cyan), both in a box $40h^{-1}$ comoving Mpc on a
  side. Also shown are the overdensity profiles for 1D spherical
  infall (long-dashed magenta) and the BBKS linear overdensity (dotted
  black). {\it Right-hand panel:} Fractional absorption excess
  $\delta_F$ relative to the mean IGM absorption, for a spectral
  window $\Delta v=2000\kms$ across the halo systemic velocities for
  the $40h^{-1}$~Mpc comoving box simulations with and without a wind,
  shown against projected impact parameter $b_\perp$ for halo masses
  $\log_{10}(M_h/M_\odot)=11.1$ (triangles; blue), 11.5 (squares;
  magenta) and $12.0-12.6$ (circles; black) at redshift $z=2.4$. For
  clarity of presentation, the points for the lowest and highest halo
  mass bins have been interpolated to slightly offset values of
  $b_\perp$. The halo velocities include a random component drawn from
  a Gaussian distribution with standard deviation $\sigma_{\rm
    H}=520\kms$. The curves (green) correspond to a 1D spherical
  infall model around haloes with masses $\log_{10}(M_h/M_\odot)=11$,
  11.5 and 12 (see text).  }
\label{fig:df_comp_planck1_40Mpc1024_qLya_PS13_infall_bcom}
\end{figure*} 

The left-hand panel of
Fig.~\ref{fig:df_comp_planck1_40Mpc1024_qLya_PS13_infall_bcom} shows
the mean spherically averaged gas overdensity around haloes of mass
$10^{12}\,\MSun$ at redshift $z = 2.4$. The profiles are constructed
by spherically averaging the gas distribution around each halo in
radial bins, and then averaging the resulting spherical profiles over
the haloes in a narrow halo mass bin centred at
$10^{12}\,M_\odot$. Results are shown for two simulations in a
$40h^{-1}$~Mpc comoving box, one with a dark matter particle mass
resolution of $4.30\times10^6\,h^{-1}\MSun$ (40-1024) and the other
with a resolution of $5.37\times10^5\,h^{-1}\MSun$ (40-2048). The
density profiles for the models agree well, and largely match the
predictions of the spherical collapse model. Deviations from the BBKS
linear theory profile become pronounced only within the inner 5~cMpc.

The absorption excess $\delta_F$ relative to the mean IGM, averaged
over a spectral window of width $\Delta v=2000\kms$, is shown in the
right-hand panel of
Fig.~\ref{fig:df_comp_planck1_40Mpc1024_qLya_PS13_infall_bcom} for the
non-wind simulation 40-1024 and the wind simulation 40-1024-ps13 at
$z=2.4$. A random line-of-sight halo velocity is allowed for, drawn
from a Gaussian distribution with dispersion $\sigma_{\rm H}=520\kms$,
typical of measurements \citep{2013ApJ...776..136P}. Results are shown
as a function of projected line-of-sight comoving impact parameter
$b_\perp$ for haloes in mass bins centred at
$\log_{10}(M_h/M_\odot)=11.1$ and 11.5 of width
$\Delta\log_{10}(M_h/M_\odot)=0.2$, and for haloes with masses $12.0 <
\log_{10}(M_h/M_\odot)=12.6$. The amount of absorption increases
weakly with halo mass. While the wind simulation produces a greater
amount of absorption for all impact parameters, $\delta_F$ is enhanced
by the wind by an amount exceeding 0.05 only within the inner 50~ckpc
for $\log_{10}(M_h/M_\odot)=11.1$, 70~ckpc for
$\log_{10}(M_h/M_\odot)=11.5$ and 200~ckpc for $12.0 <
\log_{10}(M_h/M_\odot)<12.6$. For comparison, the corresponding virial
radii are 160~ckpc, 220~ckpc and $320-510$~ckpc, respectively, so that
this level of enhancement occurs at about one-third the virial
radii. As shown in the Appendix, when the impact parameters are
rescaled to the virial radius for each halo mass, both the profiles
without wind feedback and with are each nearly universal, independent
of halo mass.

The spherically symmetric collapse models similarly predict an
increase in $\delta_F$ with halo mass, as shown by the (green) curves
in Fig.~\ref{fig:df_comp_planck1_40Mpc1024_qLya_PS13_infall_bcom}
(right-hand panel). The spherically symmetric solution is for an
isolated halo. It neglects the average enhancement in absorption
within the velocity window from the chance interception of other
haloes clustered with the central halo. Estimating the correction
analytically is not straightforward as it involves modelling the
clustering of haloes in an overdense region. Instead we allow for a
small additional amount of absorption common to all halo masses,
normalizing to the absorption in the simulations at 7~cMpc. Allowing
for the correction, the spherically symmetric model predictions agree
closely with the non-wind numerical simulation from asymptotically
large intergalactic scales to within the virial radii of the haloes.

\subsection{Comparison with observations}

\begin{figure}
\scalebox{0.5}{\includegraphics{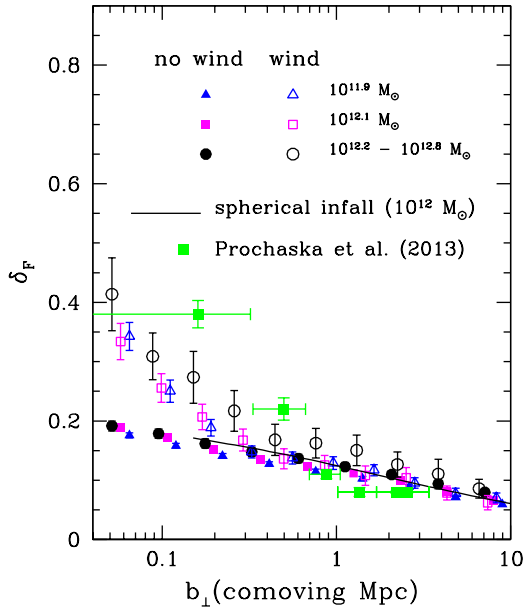}}
\vspace{-1.5cm}
\caption{Absorption signature around QSOs. The fractional absorption
  excess $\delta_F$ relative to the mean IGM absorption is shown for a
  spectral window $\Delta v=2000\kms$ across the halo systemic
  velocities for the model without a wind (80-2048; solid symbols) and
  with a wind (40-1024-ps13; open symbols) at redshift $z=2.4$, for
  halo masses $\log_{10}(M_h/M_\odot)=11.9$ (triangles; blue), 12.1
  (squares; magenta) and $12.2-12.8$ (circles; black). The solid line
  shows the prediction for simple spherical accretion. The data points
  (large green filled squares) are for absorption around QSOs at
  $z\approx2.4$ taken from \citet{2013ApJ...776..136P}, with the error
  bars showing the errors in the mean.
}
\label{fig:df_comp_planck1_80Mpc2048qLya_40Mpc1024PS13_infall_P13_bcom_RT}
\end{figure} 

We use the integrated absorption excess over broad velocity windows to
quantify the absorption around galaxy haloes. Previously we found this
provided an adequate description of the effect of overdense structures
near the haloes on the absorption statistics
\citep{2015MNRAS.453..899M}.

In
Fig.~\ref{fig:df_comp_planck1_80Mpc2048qLya_40Mpc1024PS13_infall_P13_bcom_RT}
we compare the predictions of models without and with winds to the
$\delta_F$ measurements surrounding QSOs reported by
\citet{2013ApJ...776..136P} at $z\approx2.4$, with a velocity window
$\Delta v=2000\kms$. A random component drawn from a Gaussian
distribution with $\sigma_{\rm H}=520\kms$ is added to the halo
velocities in the simulations to match the typical errors in the
measured halo redshifts. The mean QSO halo mass at this epoch is
$(1-3)\times10^{12}\MSun$ \citep{2012ApJ...752...39T,
  2012MNRAS.424..933W, 2015MNRAS.453.2779E}. Outside the virial radii
(corresponding to $300-590$ comoving kpc for the halo masses shown),
the measurements agree closely with the simulation predictions. The
simulations slightly over-predict the amount of absorption near $1-2$
comoving Mpc. This may indicate too low a photoionization background
was used, but we conservatively maintain the level used in the
simulations. The measured values also match the expectation for
spherical accretion onto a halo. The data show excess absorption over
the non-wind model predictions at a statistically significant level
(more than $3\sigma$) out to impact parameters of $300-700$ comoving
kpc, with the excess increasing with decreasing impact parameter. The
wind model shows excess absorption over the non-wind model within the
virial radii for the most massive haloes, but the amount still falls
short of the measured values. The central two data points are more
dominated by haloes with $z\simeq2.0$ \citep{2015MNRAS.452.2034R}. The
values for $\delta_F$ in the simulation at $z = 2.0$ agree within the
errors with those at $z = 2.4$ or are somewhat smaller. The results
suggest that while this particular wind model performs better than the
constant velocity wind models studied previously
\citep{2015MNRAS.453..899M}, it is still not able to account fully for
the high amounts of measured absorption. Alternatively, it may be an
indication of the need to include AGN feedback.

\begin{figure}
\scalebox{0.5}{\includegraphics{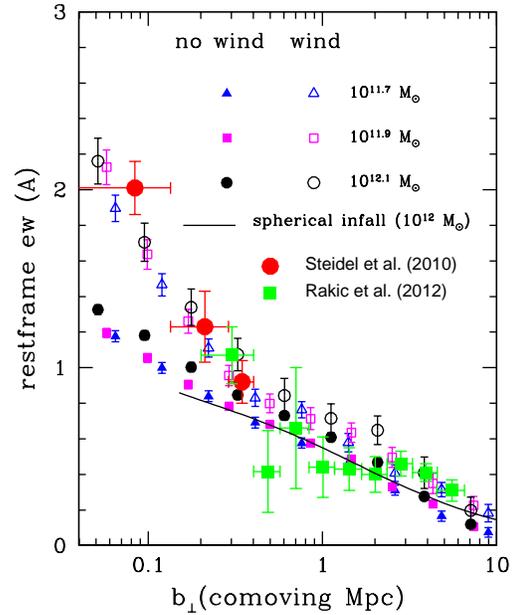}}
\vspace{-1.5cm}
\caption{Absorption signature around star-forming galaxies. The
  rest-frame equivalent width ($\rm \AA$) is shown within a velocity
  window $\Delta v=1000\kms$ centred on the halo centre-of-mass
  velocity as a function of the line-of-sight impact parameter
  $b_\perp$ for halo masses $\log_{10}(M_h/M_\odot)=11.7$ (triangles;
  blue), 11.9 (squares; magenta) and 12.1 (circles; black) at redshift
  $z=2.4$ for models without wind feedback (80-2048; solid symbols)
  and with (40-1024-ps13; open symbols). The solid line shows the
  prediction for simple spherical accretion. The data are for
  star-forming galaxies from \citet{2010ApJ...717..289S} (red filled
  circles) and \citet{2012ApJ...751...94R} (large green filled
  squares).
}
\label{fig:ew_R12}
\end{figure}

In Fig.~\ref{fig:ew_R12}, we compare with the absorption equivalent
widths measured around star-forming galaxies at $z\approx2.4$ within
$1000\kms$ wide velocity windows centred on the systemic velocities of
the galaxies using the data from \citet{2010ApJ...717..289S} (red
filled circles) and \citet{2012ApJ...751...94R} (large green filled
squares). A random component drawn from a Gaussian distribution with
$\sigma_{\rm H}=130\kms$ is added to the halo velocities in the
simulations to match the typical errors in the measured halo
redshifts. A spatial correlation analysis gives an estimated median
halo mass for the galaxies in the survey of $10^{11.9\pm0.1}\MSun$
\citep{2012ApJ...752...39T}. We show results for halo mass bins
centred at $\log_{10}(M_h/M_\odot) = 11.7$, 11.9 and 12.1 of width
$\Delta\log_{10}(M_h/M_\odot)=0.2$. Following the observational
analysis procedure of \citet{2012ApJ...751...94R} used to match their
newer data to the continuum level of the earlier data of
\citet{2010ApJ...717..289S}, the simulated spectra have been adjusted
by applying a renormalization factor of 0.804 to the fluxes. Outside
the virial radii ($250-350$ comoving kpc for the halo masses shown),
the predicted equivalent widths of all the models agree well with the
data within the measurement errors. They also agree with the
prediction for spherical infall without feedback. Excess absorption
over the non-wind model predictions is detected at a statistically
significant level (more than $4\sigma$) only within the inner
$b_\perp<140$ comoving kpc bin around the galaxies. The wind model now
boosts the amount of absorption to match the measured levels. This
contrasts with the constant velocity wind model, for which too little
absorption was found \citep{2015MNRAS.453..899M}.

\begin{figure}
\scalebox{0.5}{\includegraphics{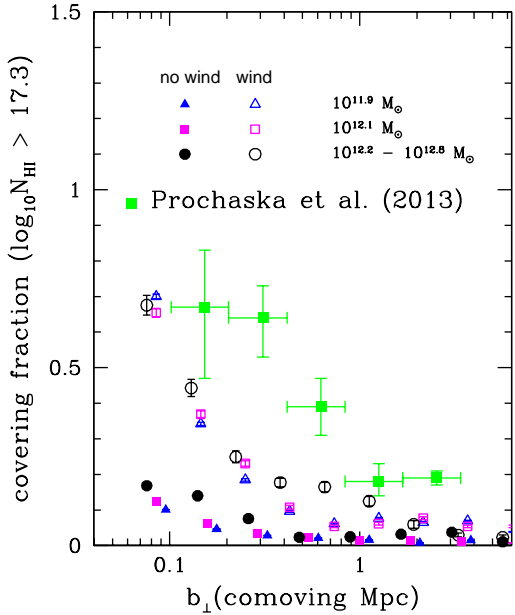}}
\vspace{-1.5cm}
\caption{Covering fraction around QSOs. The covering fraction of
  \HI\ gas with column density $\log_{10}N_{\rm HI}>17.3$ is shown for
  a spectral window $\Delta v=3000\kms$ across the halo systemic
  velocities for the model without a wind (80-2048; solid symbols) and
  with a wind (40-1024-ps13; open symbols) at $z=2.4$. The data points
  (large green filled squares) are from \citet{2013ApJ...776..136P}
  for absorption around QSOs at $z\sim2.4$, with the error bars
  showing the errors in the mean.}
\label{fig:fcov_absorbers_comp_planck1_40_2048_vs_ps13subf_P13_lgNhi17p3_logbperp}
\end{figure} 

\begin{figure}
\scalebox{0.5}{\includegraphics{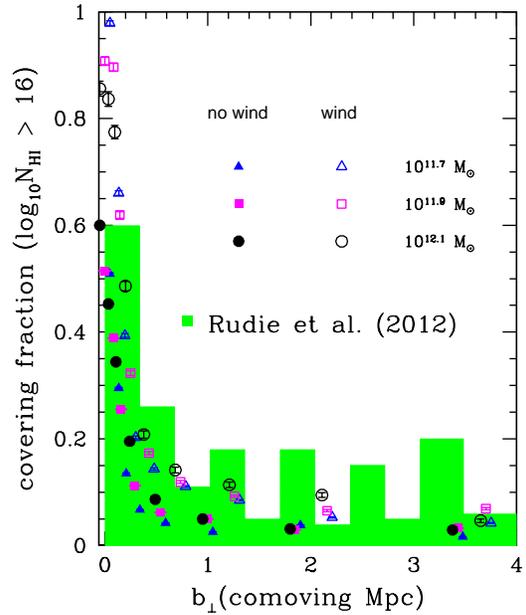}}
\vspace{-1.5cm}
\caption{Covering fraction around star-forming galaxies. The covering
  fraction of \HI\ gas with column density $\log_{10}N_{\rm HI}>16.0$
  is shown for a spectral window $\Delta v=600\kms$ across the halo
  systemic velocities for the model without a wind (80-2048; solid
  symbols) and with a wind (40-1024-ps13; open symbols) at
  $z=2.4$. The data are from \citet{2012ApJ...750...67R} for
  star-forming galaxies (green histogram).
}
\label{fig:fcov_absorbers_comp_planck1_40_2048_vs_ps13subf_lgNHI16_linbperp}
\end{figure}

We have also checked against \HI\ covering fractions around QSOs and
star-forming
galaxies. Fig.~\ref{fig:fcov_absorbers_comp_planck1_40_2048_vs_ps13subf_P13_lgNhi17p3_logbperp}
shows the fraction of lines of sight through haloes with masses
characteristic of QSOs that contain absorption systems with
\HI\ column densities exceeding $10^{17.3}\,{\rm cm^{-2}}$ at $z =
2.4$.  The predicted covering fractions at $z = 2.0$ agree within the
errors with those at $z = 2.4$ or are somewhat smaller. As we found
previously \citep{2015MNRAS.453..899M}, the covering fraction
continues to fall short of the measurements of
\citet{2013ApJ...776..136P}, even for the new supernovae-driven wind
model used here. Part of the discrepancy, such as near $1-2$ comoving
Mpc, may be attributed to unresolved absorption systems
\citep{2013ApJ...776..136P}, but not the steep rise towards small
impact parameters. By contrast,
Fig.~\ref{fig:fcov_absorbers_comp_planck1_40_2048_vs_ps13subf_lgNHI16_linbperp}
shows that the new wind model now recovers the \HI\ covering fraction
of systems around haloes with masses typical of star-forming galaxies,
as measured by \citet{2012ApJ...750...67R} for absorption \HI\ column
densities exceeding $10^{16}\,{\rm cm^{-2}}$.
\citet{2012ApJ...750...67R} also report cumulative covering fractions
of 20 per cent. for systems with column densities in the range
$10^{17.2}-10^{20.3}\,{\rm cm^{-2}}$ for impact parameters within both
the virial radius and twice the virial radius. We find 18 per
cent. and 8 per cent., respectively. While the cumulative covering
fraction within the virial radius is comparable to the measured value,
the fraction within twice the virial radius is somewhat low.

\section{Discussion and conclusions}
\label{sec:conclusions}

We use the Sherwood simulations, a suite of cosmological
$N$-body$+$hydrodynamics IGM simulations, to predict absorption
signatures around QSOs and star-forming galaxies. Both models without
and with feedback in the form of supernovae-driven winds are
included. No metals are included. In this paper, we use an
energy-driven feedback model that scales the wind velocity with the
escape velocity from the halo and varies the amount of mass loading as
its inverse square, using a constant amount of mechanical energy per
unit mass in stars formed to drive the wind. This contrasts with our
previous treatment which invoked a wind with constant velocity and
constant mass loading independent of halo mass.

A comoving box size of 40$h^{-1}$Mpc with $1024^3$ dark matter and gas
particles each is shown to provide a well-converged estimate of the
amount of absorption, when there are sufficient numbers of haloes. To
produce accurate statistics for QSO haloes, however, a larger box size
is preferred. We use a comoving box size of 80$h^{-1}$Mpc with
$2048^3$ dark matter and gas particles each for comparison of the
non-wind simulations with all observations. Because of the greater
expense of the wind simulations, a 40$h^{-1}$Mpc comoving box was
used. An approximate analytic formulation is used to estimate the
effects of radiative transfer on the absorption signatures in a
post-processing step. For models without winds, the radiative transfer
corrections negligibly affect the absorption signature, but the
effects are non-negligible when a wind is included. The radiative
transfer correction was applied in all cases for
consistency. Radiative hydrodynamics effects are not taken into
account.

Excellent agreement is found between the measured and predicted
mesogalactic neutral hydrogen absorption signatures around
star-forming galaxies outside the virial radius. Both line-of-sight
absorption measurements and neutral hydrogen covering fractions are
reproduced. This demonstrates that there is no need for the
introduction of additional physical assumptions to describe the gas
density and peculiar velocity fields for gas on these scales. Good
agreement is also found well outside the virial radius for the QSO
data. A small excess in the predicted amount of absorption may be a
result of an under-estimated photoionizing radiation field. This may
reflect an under-estimate of the metagalactic ionizing radiation
background. It may also be an indication of a transverse proximity
effect resulting from a local contribution to the ionizing radiation
from the QSOs themselves, for which there is some independent evidence
\citep{2003A&A...397..891J, 2014ApJ...784...42S}. Outside the virial
radius, the spherically averaged absorption signatures are still well
described by spherical infall of intergalactic gas onto the haloes for
both star-forming galaxies and QSOs, as predicted for the extended
haloes of galaxies and QSOs forming at the peaks of a primordial
gaussian field of density fluctuations, allowing for secondary infall.

By contrast, for impact parameters lying within the virial radii,
statistically significant excess absorption is measured for both
star-forming galaxies and QSOs compared with the non-feedback
predictions.  The excess absorption extends to substantially larger
distances, of $300-700$ comoving kpc, around haloes exceeding
$\log_{10}(M_h/M_\odot)=12$, typical of those harbouring QSOs, than
for the less massive haloes in which star-forming galaxies usually
reside, for which excess absorption is measured only within the inner
$\sim150$ comoving kpc. This suggests that the affected region is
larger around QSOs either because the feedback impact on the
surrounding gas depends on halo mass beyond the scaling with virial
radius found for the feedback model used, or because of a difference
in the feedback mechanisms for star-forming galaxies and QSOs.

The supernovae-driven wind model used accounts for both the excess
absorption around star-forming galaxies and the neutral hydrogen
covering fraction, unlike our previous wind model simulations. The
agreement suggests neither AGN heating nor metal cooling are required
to match the absorption signatures for star-forming galaxies, although
this conclusion may depend on the feedback model used. By contrast,
even for the model with feedback, the simulations fail to account for
the amounts of excess absorption and the covering fractions of
optically thick gas near QSOs, even though the feedback model predicts
the boost in absorption from the wind should increase with the halo
mass at a fixed impact parameter. Additional heating by black-hole
accretion may be required, although it may also indicate a need for
higher numerical resolution of the gas.
 
\section*{Acknowledgments}
This work used the Curie supercomputer at the Tr\'e Grand Centre de
Calcul (TGCC) in France, made available through time awarded by the
Partnership for Advanced Computing in Europe (PRACE) 8th Call.  This
work also made use of the DiRAC High Performance Computing System
(HPCS) and the COSMOS shared memory service at the University of
Cambridge. These are operated on behalf of the STFC DiRAC HPC
facility. This equipment is funded by BIS National E-infrastructure
capital grant ST/J005673/1 and STFC grants ST/H008586/1,
ST/K00333X/1. We thank V. Springel for making \texttt{P-GADGET-3}
available.  JSB acknowledges the support of a Royal Society University
Research Fellowship.  EP acknowledges support by the Kavli Foundation.

\bibliographystyle{mn2e-eprint}
\bibliography{ms}

\appendix 

\section{Convergence tests on \HI\ statistics}

\begin{figure}
\scalebox{0.5}{\includegraphics{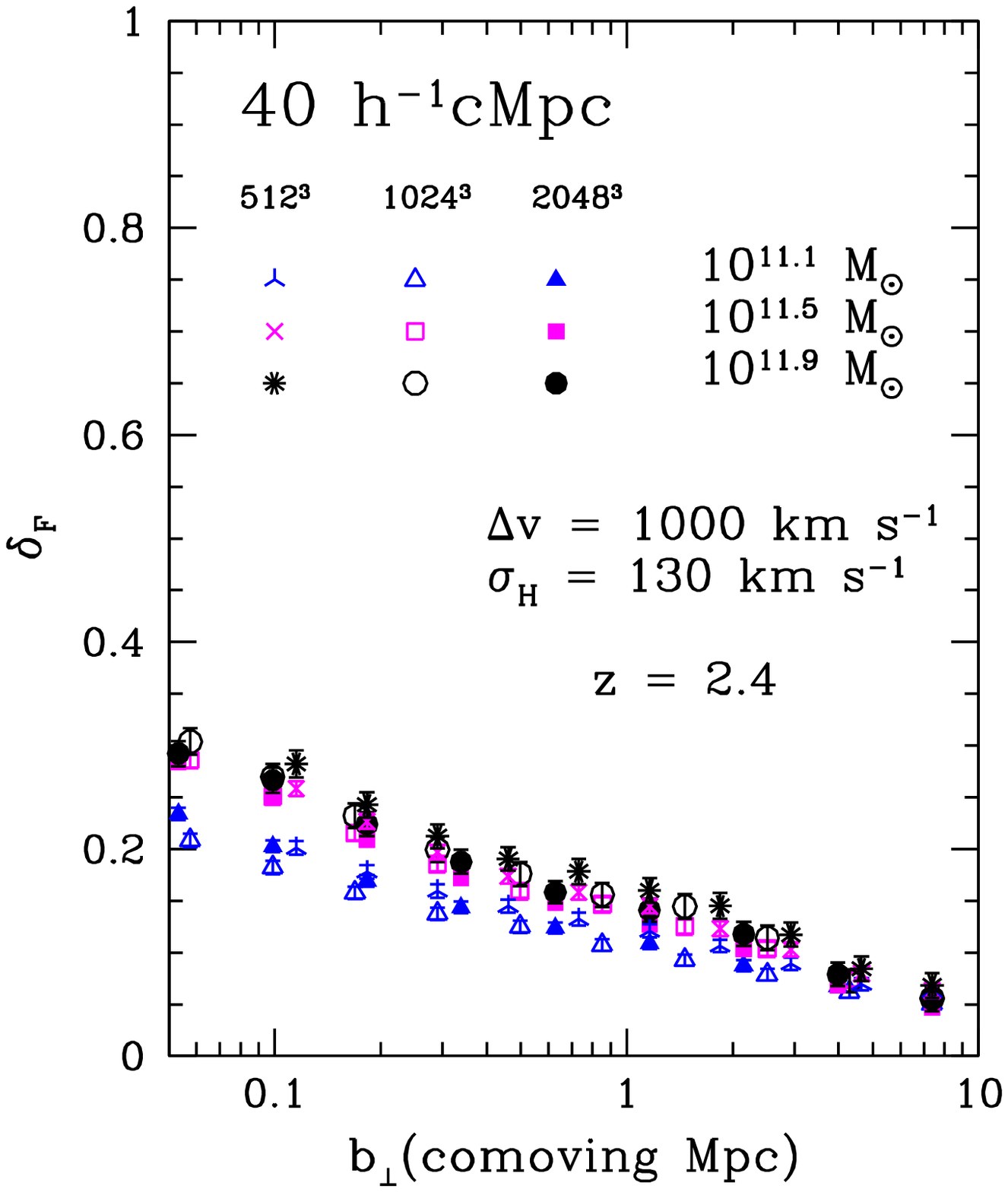}}
\vspace{-1.5cm}
\caption{Fractional absorption excess $\delta_F$ relative to the mean
  IGM absorption for a velocity window $\Delta v=1000\kms$ around the
  halo centre-of-mass velocity. The data are shown as a function of
  the line-of-sight impact parameter $b_\perp$ for halo masses
  $\log_{10}(M_h/M_\odot)=11.1$ (blue triangles),
  $\log_{10}(M_h/M_\odot)=11.5$ (magenta squares) and
  $\log_{10}(M_h/M_\odot)=11.9$ (black circles) at $z=2.4$ for
  simulations in a comoving box size 40$h^{-1}$~Mpc with dark matter
  particle mass resolutions of $3.44\times10^7\,h^{-1}\MSun$ ($512^3$
  particles; starred symbols), $4.30\times10^6\,h^{-1}\MSun$ ($1024^3$
  particles; open symbols), and $5.37\times10^5\,h^{-1}\MSun$
  ($2048^3$ particles; filled symbols). No systematic differences with
  resolution are found within the errors, although the lowest
  resolution simulations show a small tendency towards slightly larger
  absorption deficits. Halo velocities include a random component
  drawn from a Gaussian distribution with standard deviation
  $\sigma_{\rm H}=130\kms$.
}
\label{fig:df_comp_40Mpc_512_1024_2048_velw500_RT}
\end{figure}

The convergence of the fractional absorption excess $\delta_F$ at
$z=2.4$ with mass resolution at a fixed comoving box size of
$40\,h^{-1}{\rm Mpc}$ is tested in
Fig.~\ref{fig:df_comp_40Mpc_512_1024_2048_velw500_RT} for a velocity
window of width $\Delta v=1000\kms$ and a halo redshift uncertainty
$\sigma_{\rm H}=130\kms$, similar to the of observations of
\citet{2012ApJ...751...94R}. A weak trend of increasing absorption
with halo mass is found for decreasing impact parameters. For a given
mass halo, there is no significant trend with particle mass,
demonstrating the fluctuations in the gas giving rise to the
absorption are well resolved. The absorption results shown in this
paper use a mass resolution corresponding to a dark matter particle
mass of $4.3\times10^6\,h^{-1}M_\odot$.

\begin{figure}
\scalebox{0.5}{\includegraphics{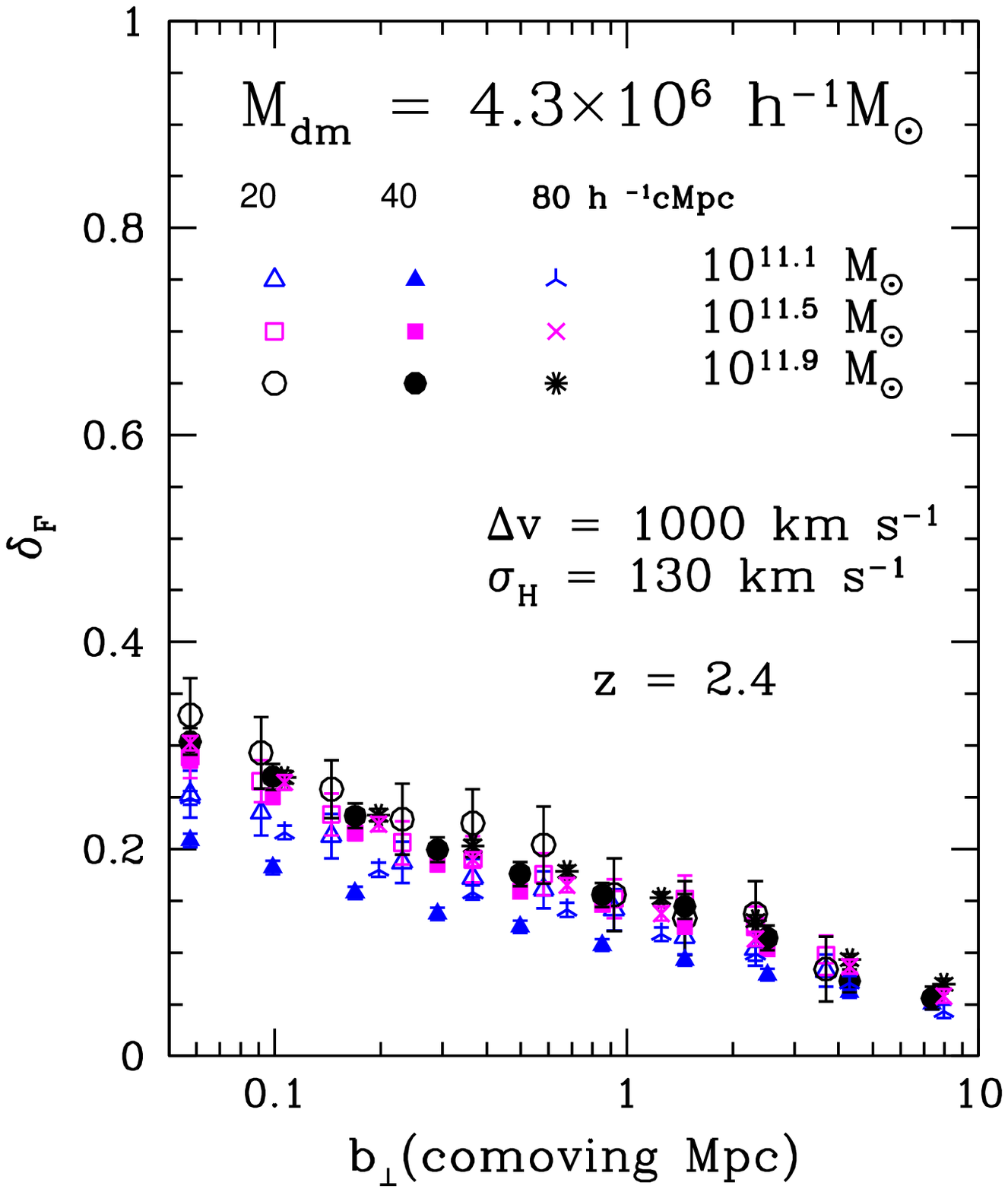}}
\vspace{-1.5cm}
\caption{Fractional absorption excess $\delta_F$ relative to the mean
  IGM absorption within a $\Delta v=1000\kms$ window centred on the
  halo centre-of-mass velocity.  The data are shown as a function of
  the line-of-sight impact parameter $b_\perp$ for halo masses
  $\log_{10}(M_h/M_\odot)=11.1$ (blue triangles), 11.5 (magenta
  squares) and 11.9 (black circles) at redshift $z=2.4$, for the
  non-wind simulations. No substantial differences are found between
  the box sizes within the errors, except for the smallest halo mass
  bin for which the convergence is somewhat slower. Halo velocities
  include a random component drawn from a Gaussian distribution with
  standard deviation $\sigma_{\rm H}=130\kms$.
}
\label{fig:df_comp_20Mpc0512_40Mpc1024_80Mpc2048_velw500_RT}
\end{figure}

The convergence of the fractional absorption excess $\delta_F$ at
$z=2.4$ with comoving box size at a fixed mass resolution,
corresponding to a dark matter particle mass $M_{\rm
  dm}\simeq4.3\times10^6\,h^{-1}M_\odot$, is tested in
Fig.~\ref{fig:df_comp_20Mpc0512_40Mpc1024_80Mpc2048_velw500_RT}. The
same trend of increasing absorption with halo mass for decreasing
impact parameter is recovered. For a given halo mass, there is no
significant trend with simulation box size, although convergence is
slowest at small impact parameters for the smallest halo mass bin.
The results show that the large-scale power giving rise to
fluctuations in the absorption has been well captured for the halo
masses of interest. The absorption results in the paper are based on a
simulation volume of comoving side $40\,h^{-1}{\rm Mpc}$ for models
with a wind and $80\,h^{-1}{\rm Mpc}$ for models without.

\begin{figure}
\scalebox{0.5}{\includegraphics{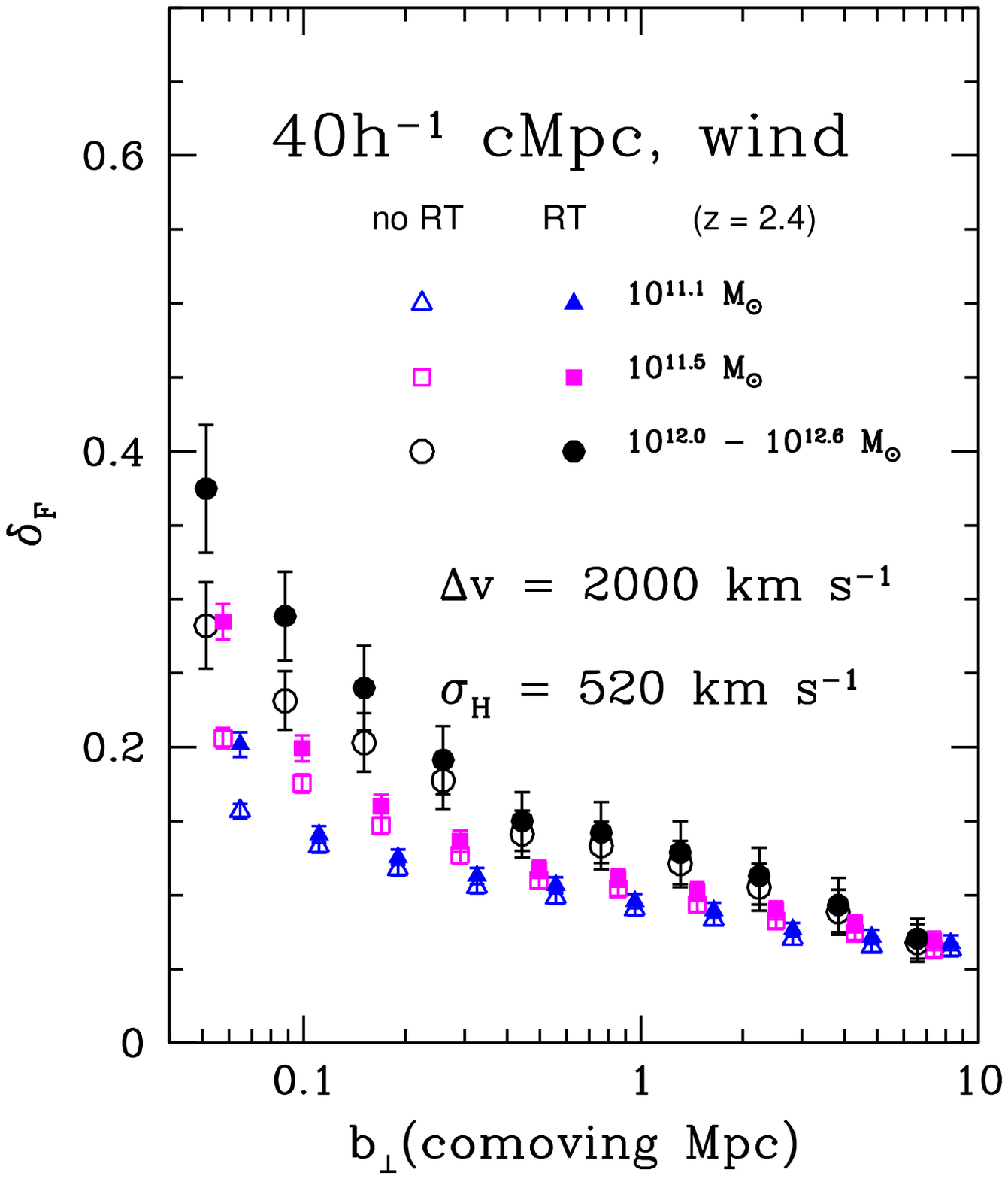}}
\vspace{-1.5cm}
\caption{Fractional absorption excess $\delta_F$ relative to the mean
  IGM for halo masses $10^{11.1}\,M_\odot$ (blue triangles),
  $10^{11.5}\,M_\odot$ (magenta squares) and
  $12.0<\log_{10}(M_h/M_\odot)<12.6$ (black circles) within velocity
  windows of width $\Delta v=2000\kms$ in a $40h^{-1}$~Mpc comoving
  box for the simulation with galactic winds, at $z=2.4$. The effects
  of corrections for radiative transfer are shown. The enhancement in
  absorption induced by the radiative transfer corrections increases
  with halo mass at a fixed impact parameter. Halo velocities include
  a random component drawn from a Gaussian distribution with standard
  deviation $\sigma_{\rm H}=520\kms$.
}
\label{fig:df_comp_planck1_40Mpc1024_PS13_bcom_noRT_RT}
\end{figure}

We examine the possible role radiative transfer may play on the amount
of absorption using the simplifying approximation of an attenuated
radiation field within systems sufficiently dense to be self-shielded
from an external photoionizing radiation field. We adopt the
semi-analytic prescription of \citet{2013MNRAS.430.2427R}, using a
characteristic self-shielding total hydrogen density of
$0.0064T_4^{0.17}\,{\rm cm^{-3}}$ for temperature $T_4=T/10^4\,{\rm
  K}$. Fig.~\ref{fig:df_comp_planck1_40Mpc1024_PS13_bcom_noRT_RT}
shows the absorption excesses in a velocity window of width $\Delta
v=2000\kms$, allowing for a halo redshift uncertainty $\sigma_{\rm
  H}=520\kms$, similar to the observations of
\citet{2013ApJ...776..136P}. For the non-wind models, we find the
effect on the mean values of $\delta_F$ is under 5 per cent., and
generally less than one per cent. By contrast, the wind simulation
shows a marked increase in $\delta_F$ when the radiative transfer
correction is included. The difference increases with increasing halo
mass and decreasing impact parameter, reaching correction factors as
high as 10--20 per cent. For consistency, in this paper we include the
effects of radiative transfer for all the model predictions of the
integrated amount of absorption.

\begin{figure}
\scalebox{0.5}{\includegraphics{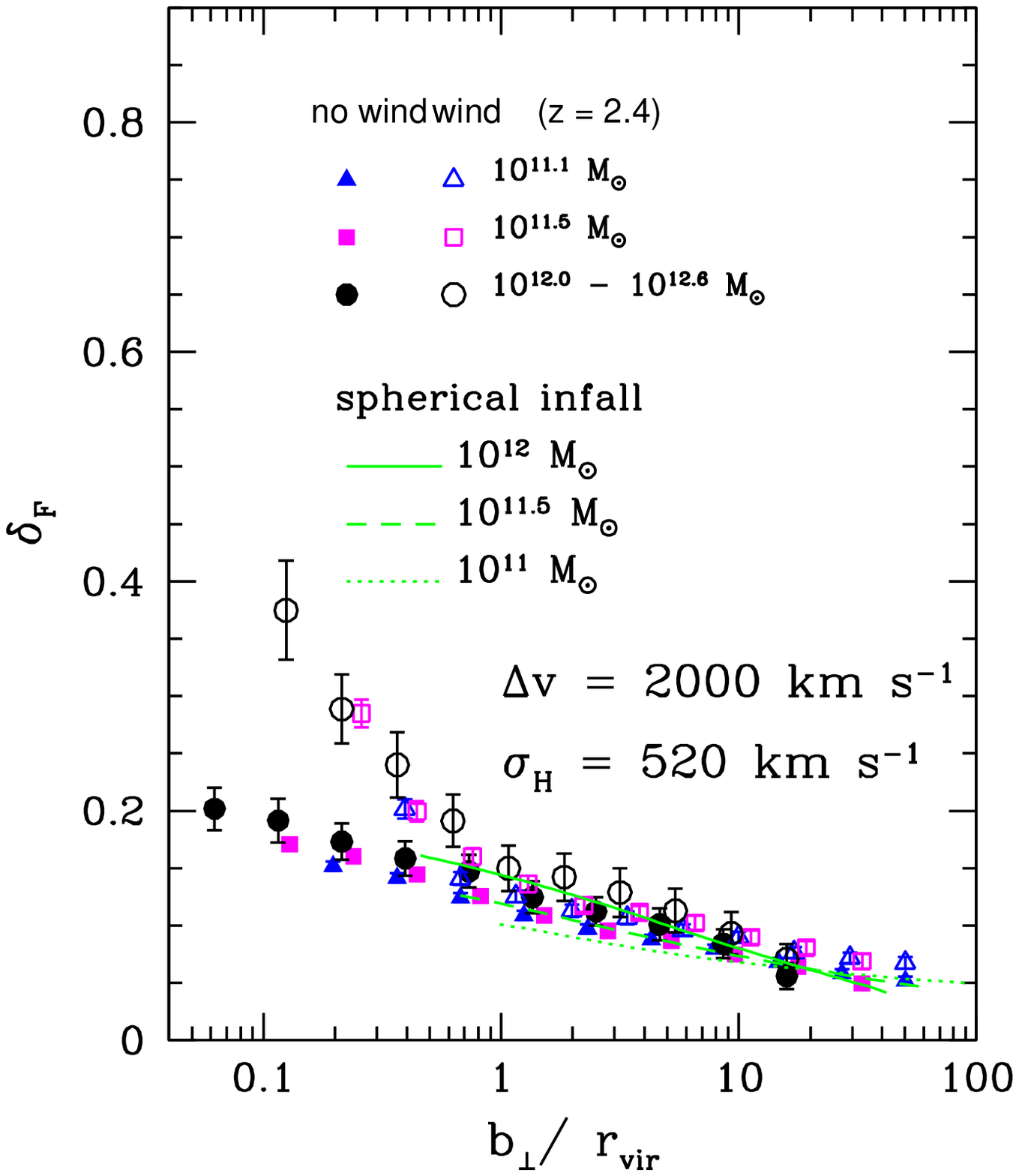}}
\vspace{-1.5cm}
\caption{Fractional absorption excess $\delta_F$ relative to the mean
  IGM absorption, for a spectral window $\Delta v=2000\kms$ across the
  halo systemic velocities for the $40h^{-1}$~Mpc comoving box
  simulations with and without a wind, shown against projected impact
  parameter $b_\perp$, scaled to the virial radius $r_{\rm vir}$, for
  halo masses $\log_{10}(M_h/M_\odot)=11.1$ (triangles; blue), 11.5
  (squares; magenta) and $12.0-12.6$ (circles; black) at redshift
  $z=2.4$. For clarity of presentation, the points for the lowest and
  highest halo mass bins have been interpolated to slightly offset
  values of $b_\perp/ r_{\rm vir}$. The halo velocities include a
  random component drawn from a Gaussian distribution with standard
  deviation $\sigma_{\rm H}=520\kms$. The curves (green) correspond to
  a 1D spherical infall model around haloes with masses
  $\log_{10}(M_h/M_\odot)=11$, 11.5 and 12.
}
\label{fig:df_comp_planck1_40Mpc1024_qLya_PS13_infall_bscaled.eps}
\end{figure} 

For the wind model, which also creates stars following a more
physically motivated model than the quick Lyman-$\alpha$ method does,
we recomputed the halo centres and peculiar velocities by centring on
the minimum of the gravitational potential rather than the
centre-of-mass. The effects on the integrated absorption statistics
and covering fractions were negligible, well within the dispersion in
the mean values, except for a small boost in the \HI\ covering
fractions for the QSOs within the innermost 800 comoving kpc, but
still within the statistical errors.

As shown in
Fig.~\ref{fig:df_comp_planck1_40Mpc1024_qLya_PS13_infall_bscaled.eps},
rescaling the impact parameter to the virial radius of the halo masses
in each mass bin results in nearly universal profiles for both the
cases without and with wind feedback, independent of halo mass to
within the Poisson errors. The enhancement in the amount of absorption
produced by wind feedback is nearly independent of halo mass when
expressed as a function of the rescaled impact parameters.

\label{lastpage}

\end{document}